# Visualization of defect-induced excitonic properties of the edges and grain boundaries in synthesized monolayer molybdenum disulfide

A. E. Yore[⊥], K.K.H. Smithe[¶], W. Crumrine[⊥], A. Miller[⊥], J. A. Tuck[⊥], B. Redd[⊥], E. Pop[¶], Bin Wang[§], A.K.M. Newaz[⊥]

[⊥]Department of Physics and Astronomy, San Francisco State University, San Francisco, CA-94132, USA
[¶]Department of Electrical Engineering, Stanford University, Stanford, CA 94305, USA
[§]School of Chemical, Biological and Materials Engineering, University of Oklahoma, Norman, OK 73019, USA

## Abstract

Atomically thin two-dimensional (2D) transition metal dichalcogenides (TMDCs) are attractive materials for next generation nanoscale optoelectronic applications. Understanding nanoscale optical behavior of the edges and grain boundaries of synthetically grown TMDCs is vital for optimizing their optoelectronic properties. Elucidating the nanoscale optical properties of 2D materials through far-field optical microscopy requires a diffraction-limited optical beam diameter sub-micron in size. Here we present our experimental work on spatial photoluminescence (PL) scanning of large size ($\geq 50$ μm) monolayer $MoS_2$ grown by chemical vapor deposition (CVD) using a diffraction limited blue laser beam spot (wavelength 405 nm) with a beam diameter as small as ~200 nm allowing us to probe nanoscale excitonic phenomena which was not observed before. We have found several important features: (i) there exists a sub-micron width strip (~500 nm) along the edges that fluoresces ~1000% brighter than the region far inside; (ii) there is another brighter wide region consisting of parallel fluorescing lines ending at the corners of the zig-zag peripheral edges; (iii) there is a giant blue shifted *A*-excitonic peak, as large as ~120 meV, in the PL spectra from the edges. Using density functional theory calculations, we attribute this giant blue shift to the adsorption of oxygen dimers at the edges, which reduces the excitonic binding energy. Our results not only shed light on defect-induced excitonic properties, but also offer an attractive route to tailor optical properties at the TMDC edges through defect engineering.

## INTRODUCTIONS

The direct band gap properties of monolayer transition metal dichalcogenides (TMDC) provide the tantalizing prospect of miniaturizing semiconductor devices to truly atomic scales and accelerating the advances of many two-dimensional (2D) optoelectronic devices.[1] 2D confinement, direct band-gap nature[2] and weak screening of charge carriers enhance the light-matter interactions[2-4] in these materials, leading to extraordinarily high absorption, electron-hole (*e-h*) creation and exciton formation (a hydrogenic entity made of an *e-h* pair). These extraordinary properties make TMDCs very attractive for optoelectronic applications[1, 5-9] including low power transistors,[10] sensitive photodetectors,[5, 11-13] energy harvesting devices,[14-16] atomically thin LEDs,[6, 8, 9] single photon sources[17-21] and nanocavity lasers.[22] To realize TMDCs for practical application, we need to understand the optical properties of edges, which become dominant as devices shrink to nanoscale.[23] The edges are also important for photocatalytic, electrocatalytic and hydrosulfphurization process[24].

In addition to the atomic edges, other structural defects may also have pronounced effects. Recent advances in chemical vapor deposition (CVD) have allowed batch production of monolayer TMDCs with macroscopic size and uniform atomic thickness.[25-28] However, these synthesized films are polycrystalline in nature largely because of the coalescence of disoriented domains,[29-31] which affect their electrical[32, 33] and optical properties.[30] It is therefore of paramount importance to understand the optical properties of the edges of these misoriented domains. So far the far-field optical properties of TMDCs have been studied using conventional scanning micro-photoluminescence (µPL) microscopy that employs micron sized or larger optical beams[29, 30, 34] lacking sufficient resolution to probe the optical properties of the edges and the grain boundaries.

Here we report our study using a scanning far-field PL mapping employing an ultra-narrow optical beam (~200 nm) that has allowed us to observe three important excitonic features at the atomic edges and the grain boundaries of large size chemical vapor deposited (CVD) grown large monolayer-$MoS_2$ (1L-$MoS_2$) flakes. First, there exists a submicron (width ~500 nm) narrow strip of higher fluorescent region at the outermost edges of monolayer $MoS_2$ flakes. Second, there is a wide (~5-10 µm) fluorescent region just inside the peripheral edge, comprised of multiple parallel submicron fluorescent lines. Finally, the *A*-exciton originating from both the single crystalline edges and polycrystalline boundaries is significantly blue shifted (~120 meV) relative to interior fluorescence. Through density functional theory (DFT) calculations, we attribute the giant blue shift phenomenon of the *A*-exciton to the reduced screening strength originating from the adsorption of oxygen dimers at the edges and grain boundaries. Our study not only elucidates the intricate defect-related luminescence properties in the edges, but also provides a pathway, through defect engineering, for the next generation of optical and optoelectronic devices.

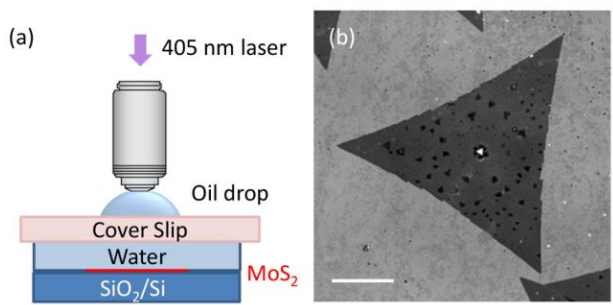

**Figure 1:** (a) The experimental setup is shown schematically (not to scale). The sample was covered by water and encapsulated by a cover slip (thickness~0.17 mm). We have used an oil objective to enhance the numerical aperture. (b) The reflection image of a 1L-$MoS_2$ sample. The scale bar is 20 µm. Note that the darker specs are small multilayer nucleation flakes.

## RESULTS AND DISCUSSION

We conducted high resolution PL scanning using a laser scanning confocal microscope (Zeiss 710 Axiovert) with an oil objective of high numerical aperture (NA=1.4). Here we raster scanned a focused excitation laser (λ~ 405 nm) over the sample and recorded PL spectra at each point. The sample was covered by a thin film of water, encapsulated by a cover slip, as shown schematically in Figure 1a. We successfully focused the beam to a spot size of diameter ~200 nm, i.e. the beam area is ~25 times smaller than the micron sized beam area used in conventional scanning PL system. The beam diameter was confirmed with a calibrated fluorescent bead of diameter 200 nm (see Supporting Information), and all measurements were conducted at room temperature. The power intensity of the laser was ~3 mW and the dwell time at each pixel~100 µs. The reflection image of one of the 1L-$MoS_2$ sample is shown in Figure 1b. The darker region on the flake is originating from bilayer spots. We have studied in total 11 samples and all demonstrate similar results (see Supporting Information). Since this high resolution microscopy requires a thin aqueous layer between a cover slip and the sample, all measurements were carried out in air at room temperature.

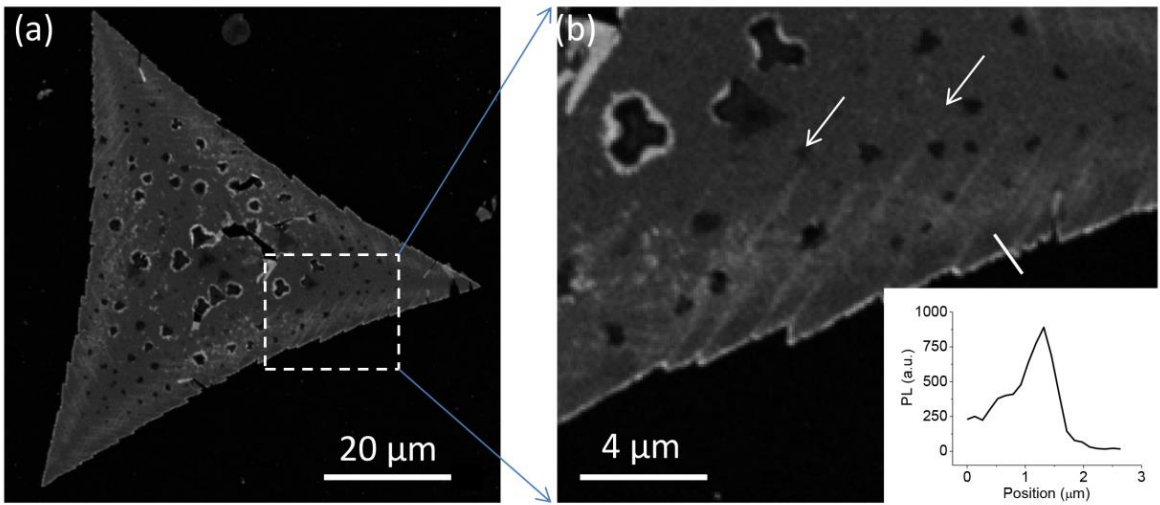

**Figure 2:** (a) The fluorescence image of a 1L-MoS$_2$ flake. (b) Blow-up view of the dashed-square in (a). The fluorescent image is showing many parallel brighter lines matching with the zig-zag edges and marked by white arrows. The inset is showing the intensity profile along the white line drawn on the edges.

High resolution fluorescence image of a large triangular 1L-MoS$_2$ sample is shown in Figure 2a. In all the previous studies, the researchers have reported a wide region of fluorescence at the edges.[30, 35] What is unique in our high resolution spatial mapping is that we successfully resolved the fluorescing regions near the edges and found that the brighter edges actually consists of two distinct regions of brighter intensity.

First, our spatial mapping shows a submicron fluorescent strip of width ~500 nm along the flake edge. The intensity profile along a line drawn across this strip is shown inset in Figure 2b, and reveals fluorescence intensities of ~400% higher than inner regions. For three samples, we observed edge fluorescence intensity ~1000% larger than interior (see Supporting Information). This extraordinarily large spike in intensity may arise from point defects, which can trap free charge carriers and localized excitons.[36] Using scanning tunneling microscopy (STM), it has been demonstrated that point defects, such as S vacancies, could form at flake edges.[37, 38]

Second, our spatial PL mapping shows another wider band along the edges consisting of parallel fluorescent lines(as marked by arrows in Figure 2b). Interestingly, these parallel lines originate from the zig-zag corners of the periphery, as is clearly visible in the blow-up image presented in Figure 2b. These fluorescent parallel lines have a spatial width of ~500nm and fluoresce with intensity ~100% higher than neighboring spots. We suggest that these fluorescent lines originate from line defects, which are boundaries between two opposite rotation domains of the 3-fold symmetric MoS$_2$ lattice, as has been observed in STM measurements of physical vapor deposited MoS$_2$ on Au (111) substrat,[39] and scanning transmission electron microscopy (TEM) of CVD-grown MoS$_2$ on SiO$_2$.[37] Here we have observed the *optical* signature of these line defects for the first time. This suggests that, although the reflection image of a large 1L-MoS$_2$ shows seamless integration into a single crystalline entity (as in Figure 1b), the flake actually consists of multiple domains, as evidenced by the fluorescing line defects along grain boundaries. Brighter luminescence from the defects suggests that these line defects may trap excitons and similar to the point defects at the periphery.

To understand the origin of the brighter edges, we have further investigated wavelength-resolved PL scans of our 1L-MoS$_2$ samples. After pumping the flakes with a ~405 nm laser, we observed a major luminescence peak shift across the sample, ranging from 660 nm at the edge to 710 nm further inwards. This peak in PL, also known as *A*-exciton peak (A-peak), has been attributed to the recombination of photoexcited excitons across the direct band gap at the K-point.[2, 4] The spatial mapping of the wavelength values of the *A*-peak for that same large triangular flake and another polycrystalline sample are shown in Figure 3a and Figure 3c-d, respectively. Figure 3b shows the PL spectra of two locations in the sample shown in Figure 3a, one at the periphery, the other just inside. Here the peak positions were obtained by fitting the PL data for every pixel using a Gaussian profile. For the pixels for which Gaussian fitting was not possible, i.e., for the pixel outside the flake, the color was set to white. A very strong blue shifted *A*-peak is observed at the flake edges relative to interior. The maximum energy of the edge *A*-peak is ~1.89 eV (~656 nm) and the minimum energy of the interior *A*-peak is 1.77 eV (~703 nm). This is a total upshift of ~120 meV, approximately ~3 times larger than previously reported values[35].

To investigate the effect of the beam diameter on the PL mapping, we varied laser beam size. We observed the upshift values decrease significantly as we employed a wider beam spot; specifically, the blue shift decreased to ~ 20 meV as we doubled the beam diameter (from 200 to 400 nm). For a few micron sized beam, there was no observable blue shift. These findings are expected, as a wider beam includes PL signal from areas further into the flake's interior, averaging out the *A*-Peak and minimizing the blue shift.

We will now discuss possible origins of our observations of giant blue shifts of *A*-exciton peaks. Various studies have shown that large shifts of the *A*-exciton peak can originate from edge stress or compression,[40-43] neutral exction-trion population[35], spatial confinement[44] or defects.[45-47] Specifically, it has been experimentally demonstrated that compressive strain can lead to a blue shift in the *A*-exciton peak;[43] and furthermore, DFT calculations have predicted that the edges of $MoS_2$ flakes are in fact naturally compressed relative to the inside.[48] Hence, the presence of compressive strain at the edges can cause a blue shift in the *A*-peak. To examine the effects of edge strain on our observed optical properties, we also studied the spatial scanning of PL for several 1L-$MoS_2$ samples prepared by microexfoliation, and did not observe any blue-shifted edges as shown in Fig.3e-f. We have examined more than five microexfoliated 1L-$MoS_2$ samples and observed similar results. Thus, we argue, the observed giant blue shift of the *A*-peak cannot be due to compressive strain at the edges. Observed giant blue shift of the *A*-exciton can also arise from the competition between trion and neutral exciton population. Since we have observed much larger blue shift than the binding energy of the trions (~20 meV), we can safely rule out that the observed giant blue shift is originating from the relative completion between trion and neutral exciton population.

It has been demonstrated that spatial confinement reduces screening strength and increases the Coulomb interaction and exciton binding energy.[49, 50] Hence, it is expected that spatial confinement at the edges of 1L-$MoS_2$ would modify the exciton binding energy and luminescing photon energy. The width of region for which the *A*-peak has been blue shifted, however, is ~ 500 nm, very large compared to the physical size of the excitons (~1-3 nm).[53] Moreover, we have not observed any blue shift for the A-exciton in the microexfoliated samples, where excitons will encounter similar edge confinement. Thus, it is unlikely that the large blue shift originates from the spatial confinement at the edge.

Now we will concentrate on edge defects, and how they can modify the PL *A*-peak as well as fluorescence intensity. Edge defects were explored using first-principles DFT calculations (See detailed in Methods). To model the flakes used in the experiments, here we used a nano ribbon that is periodic in one direction with a finite width in the other. All the calculations have been performed using a $MoS_2$ nanoribbon with a width of 5.2 nm. The width is long enough to eliminate the effect of the edges on the middle of the ribbon, as shown by the converged DOS when moving from the edge towards the middle in Figure 4b (top panel).

The $MoS_2$ nanoribbons used in our calculations are shown in Figure 4a and 4c. The ribbon was terminated by two edges, the S-edge and the Mo-edge. Previous studies of $MoS_2$ nanocrystals using atomic resolution STM have shown that both edges may be present.[38, 51, 52] The Mo edge exhibits a specific metallic electronic structure and has been shown to dominate in triangular $MoS_2$ nanocrystals supported on Au(111) surface.[53] The Mo edge has been shown to be the key active site for hydrogen evolution reaction[54] and hydrodesulfurization reaction.[55] Figure 4b(top) shows the projected density of states (PDOS) of each Mo atom across the nanoribbons since the *d* states of the Mo atoms dominate the band edge states. The two edges show some metallic character in agreement with a previous STM study.[53] When moving towards the middle of the ribbon, there is less difference between the Mo atoms, all of which converge to the bulk electronic structure.

It has been shown experimentally that S vacancies form at the edges,[37, 38] which can be saturated by oxygen in the current fabrication process. The oxidation of the edges has also been reported experimentally in STM measurements of air-exposed monolayer $WSe_2$.[56] Previous calculations suggested local strain at the grain boundaries in a 1L layered materials might also attract and accumulate oxygen.[57] Here we studied oxygen monomers at different locations across the nanoribbon, and found no significant change of the formation energy of a substitutional oxygen impurity (that is, replacement of an S atom by an O). However, for the oxygen dimer, which forms by adsorption of an oxygen molecule at a sulfur vacancy, it is more energetically favorable by 0.6 eV at the Mo-edge and 0.4 eV at the S-edge as compared to the middle of the ribbon (Figure 4c), which indicates a non-homogeneous distribution of oxygen dimers. Figure 4b (lower) plots the PDOS of each Mo atom when an oxygen dimer is located at the Mo-edge (the most stable configuration). A more pronounced electronic resonance from the edge atom (index $n$ = 18) is caused by the oxygen dimer, which introduces more valence electrons than the replaced S atom that previously occupied the position. In addition, there is also a gradual downshift of the valence band with respect to the Fermi level when moving towards the Mo-edge as schematically shown in Figure 4d. This downshift of the valence bands is caused by a surface dipole resulting from the oxygen-induced charge redistribution, which is also shown by the change of the local electrostatic potential plotted above the MoS$_2$ surface (see Figure S5 in Supplementary Information). The correlation between the increased electron density, which is mainly localized at the edge of the ribbon, suggested by the calculations and the blue shift of the PL at the edges found experimentally suggests that enhanced screening of the excited electron from its counterpart – the remaining hole in the valence band – may lead to a reduced exciton binding energy, and provides an explanation for the blue shift of the PL peak at the edges.

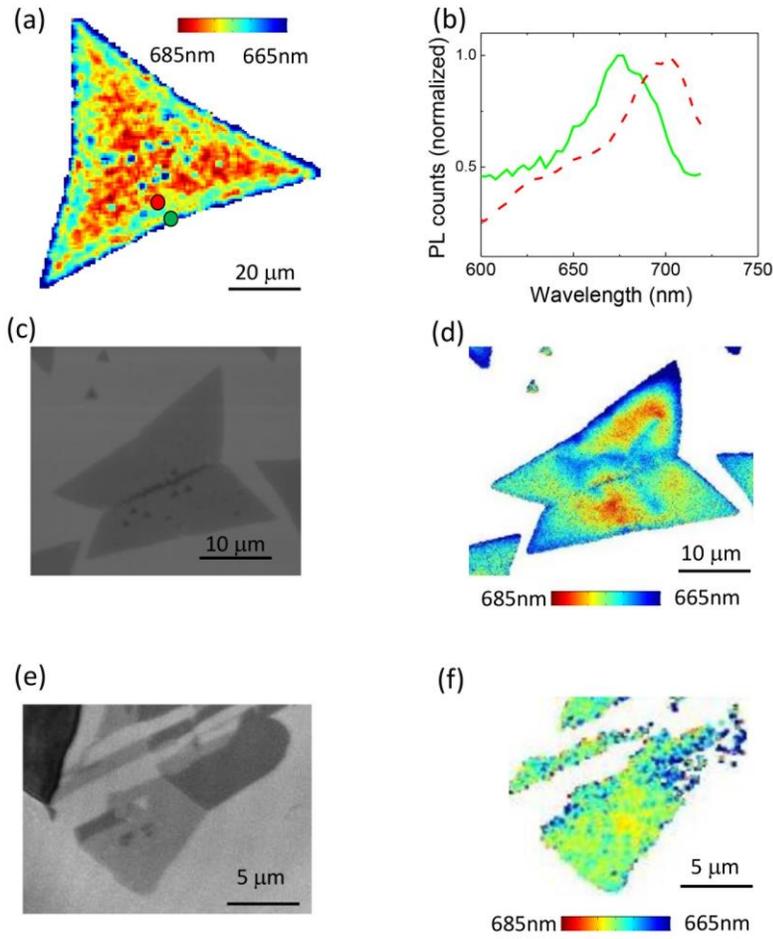

Figure 3: Spatial mapping of the A-peak position. (a) A triangular sample whose reflection image is shown in Figure 2b. (b) The red (dashed) and green color of the photoluminescence spectrum corresponds to the PL from the position marked by green and red circles, respectively, in Figure 3a. (c) Reflection image of a mirror symmetric sample. (d) Spatial mapping of the A-peak of the mirror symmetric sample. (e) Reflection image of an exfoliated monolayer (lightest contrast) connected to multilayer flakes. (f) Spatial mapping of the A-peak of the micro-exfoliated sample shown in (e).

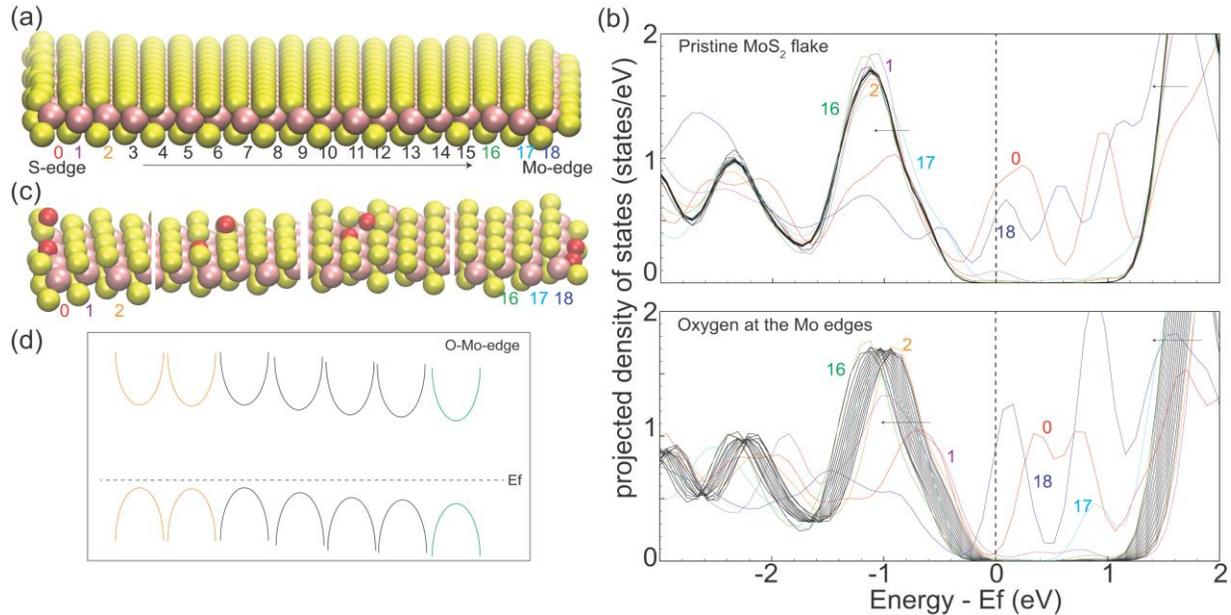

Figure 4: DFT calculations of pristine MoS$_2$ nanoribbon and oxygen functionalized MoS$_2$ nanoribbon. (a) The optimized structure and (b)-The upper and lower plots are showing the projected density of states of each Mo atoms, which are indexed in (a). (c) Different configurations of oxygen dimer at various locations in the nanoribbon, and (d) schematics to show the enhanced n-type doping caused by the oxygen dimer at the Mo-edge.

Intriguingly, our findings using a far-field optical microscope involving a ultra-narrow beam is very different from results recently reported by Bao *et al.*[35] using a ~60 nm diameter beam in a near-field scanning optical microscope setup. First, we have observed larger (~2.5 times) blue shift of the A-excitons at the periphery. Second, we have observed exciton-enhancing at the periphery in agreement to other studies[30], whereas Bao *et al.*, observed exciton-quenching luminescence. Though not understood currently, the discrepancies may be attributed to different sample growth procedures and different sizes of the samples (our samples are at least one order larger), both of which may affect the defect concentration and in turn the PL measurement.

**CONCLUSIONS**

In summary, we obtained ultra-high resolution fluorescence imaging and spatial mapping of 1L-MoS$_2$ using a narrow optical beam and found several important characteristics at the edges. We observed a strip of brighter fluorescence at the outermost edge, and another wide region of brighter luminescence consisting of submicron parallel lines originating from sharp zig-zig corners due to line defects. The latter suggests that large flakes are in fact polycrystalline at the edges. Moreover, we observed a giant blue shifted *A*-exciton peak at the edges. Through DFT calculations, we have attributed this blue shift to the oxygen dimer absorption at the edges, which reduces the binding energy of the exciton, resulting in a blue shifted *A*-exciton peak. Our results are important for informing the design of next generation truly nanoscale or atomic scale optical and optoelectronic devices, since edge effects become dominant as devices reduce in size to the nanoscale.

## MATERIALS AND METHODS:

**Sample growth:** Large flakes of 1L-MoS$_2$ were synthesized from solid S and MoO$_3$ precursors directly onto SiO$_2$ on Si, utilizing a seed layer of perylene-3,4,9,10 tetracarboxylic acid tetrapotassium salt (PTAS).[27, 58] We utilized elevated temperature (850 ºC) and atmospheric pressure to encourage epitaxial growth, which can result in single-crystal domains in excess of 200 μm on an edge (see Figure S4 in Supplementary Information). The layer thickness of the grown sample was confirmed by Raman spectroscopy[59-61] and atomic force microscopy (see supplementary information).[27, 62]

**Computational Methods:** The density functional theory (DFT) calculations were carried out using the VASP package.[63] The Perdew-Burke-Ernzerhof (PBE) generalized gradient approximation (GGA) exchange-correlation potential[64] was used, and the electron-core interactions were treated in the projector augmented wave (PAW) method.[65, 66] All the calculations have been performed using a MoS$_2$ nanoribbon with a length of 5.2 nm. Structures have been optimized using a single Γ point of the Brillouin zone with a kinetic cut off energy of 400 eV. The structure was fully relaxed until atomic forces were smaller than 0.02 eV Å$^{-1}$.

## ASSOCIATED CONTENT

**Supporting Information**

Additional information includes the results from other samples, AFM images, Raman data, the optical beam diameter measurement and supporting DFT calculations. The Supporting Information is available free of charge on the ACS Publications website.

## ACKNOWLEDGEMENT:

We thank Dr. Annette Chan for the help in measuring the spatial PL scanning. AKMN and AEY are grateful for the financial support from SFSU. KKHS and EP acknowledge support from the AFOSR grant FA9550-14-1-0251 and NSF EFRI 2-DARE grant 1542883. KKHS also acknowledges partial support from the Stanford Graduate Fellowship program and NSF Graduate Research Fellowship under Grant No. DGE-114747. This research used computational resources of the National Energy Research Scientific Computing Center, a DOE Office of Science User Facility supported by the Office of Science of the U.S. Department of Energy.

TOC

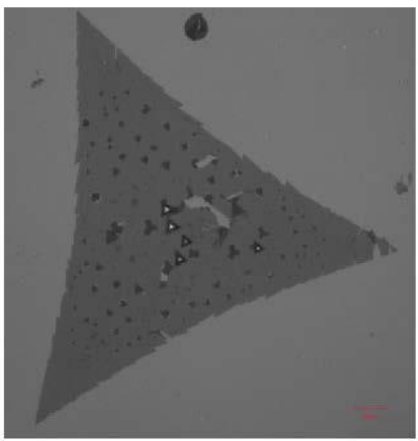
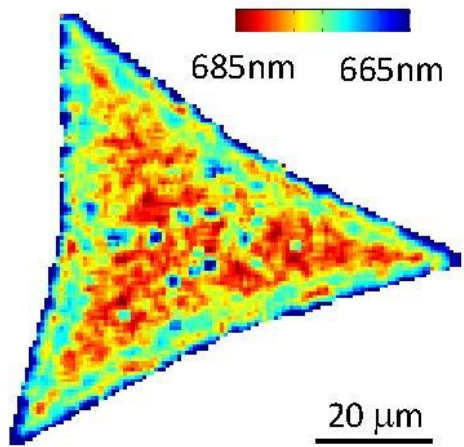

# Supplementary Information

# Defect-induced excitonic properties of the edges and grain boundaries in synthesized monolayer molybdenum disulfide

A. E. Yore[⊥], K.K.H. Smithe[¶], W. Crumrine[⊥], A. Miller[⊥], J. A. Tuck[⊥], B. Redd[⊥], E. Pop[¶], Bin Wang[§], A.K.M. Newaz[⊥]

[⊥]Department of Physics and Astronomy, San Francisco State University, San Francisco, CA-94132, USA

[¶]Department of Electrical Engineering, Stanford University, Stanford, CA 94305, USA

[§]School of Chemical, Biological and Materials Engineering, University of Oklahoma, Norman, OK 73019, USA

## S1: Blue shift of the *A*-peak and the edge fluorescent brightness

Figure S1 presents the blue shift of the *A*-exciton peak and fluorescence brightness of the edges compared to middle of the flake for all the samples.

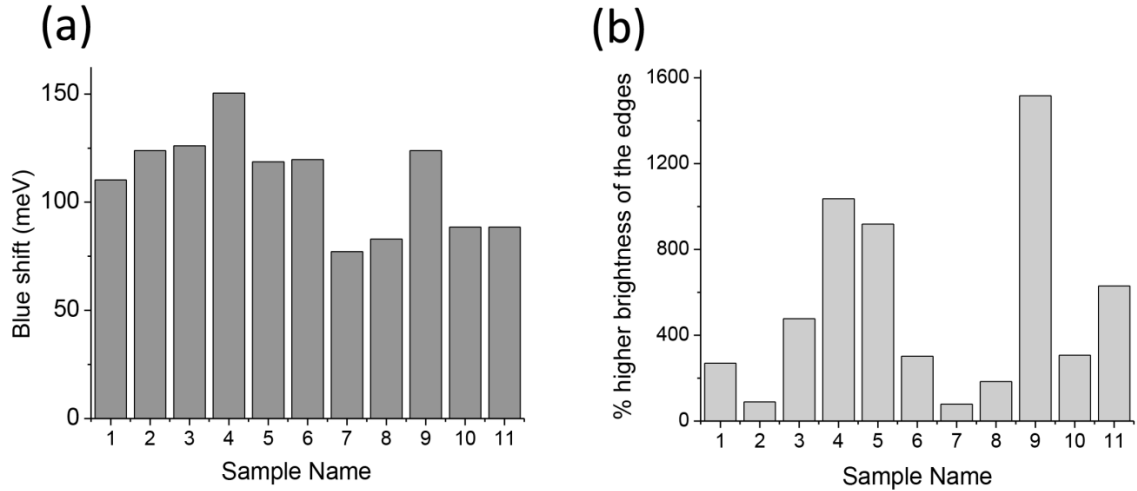

**Figure S1:** (a) The measured blue shift of the *A*-exciton peak for the PL from the edges of different samples when compared to the point inside. (b) Plot for the luminescence brightness of the outermost edges when compared to the luminescence from the point inside.

## S2: Experimental determination of the beam diameter

We have used fluorescent microspheres beads of size 200nm (Thermofisher, Catalog No-T14792). The fluorescence image of such beads is shown in Figure S2. Here the excitation laser wavelength is 405nm.

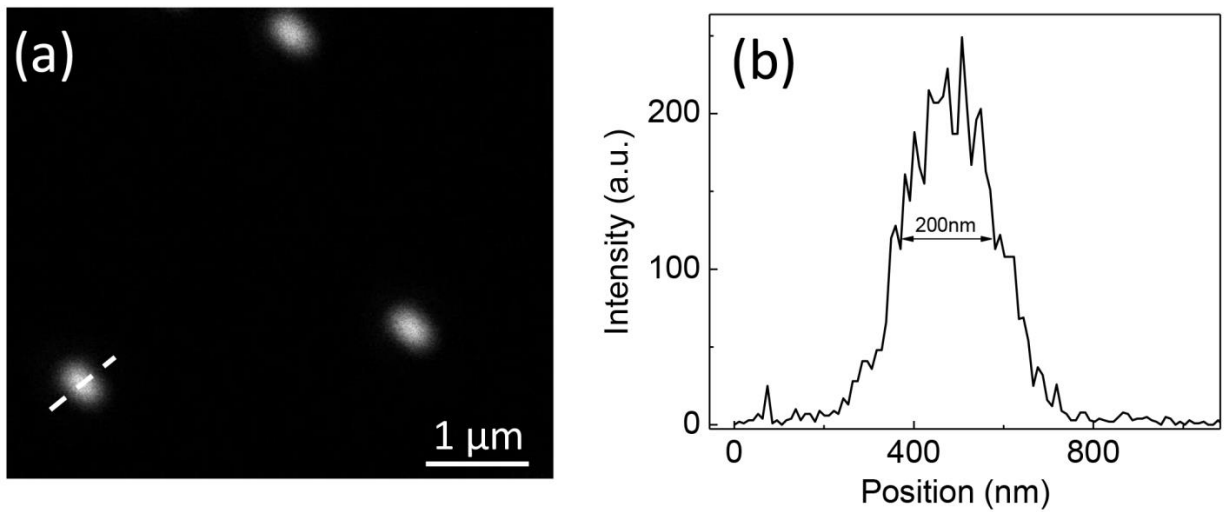

**Figure S2:** (a) Fluorescence image of microsphere beads. The scale bar is 1 micron. (b) The intensity plot along the white dashed line shown in figure (a).

## S3: Raman and AFM characterization of 1L-MoS$_2$

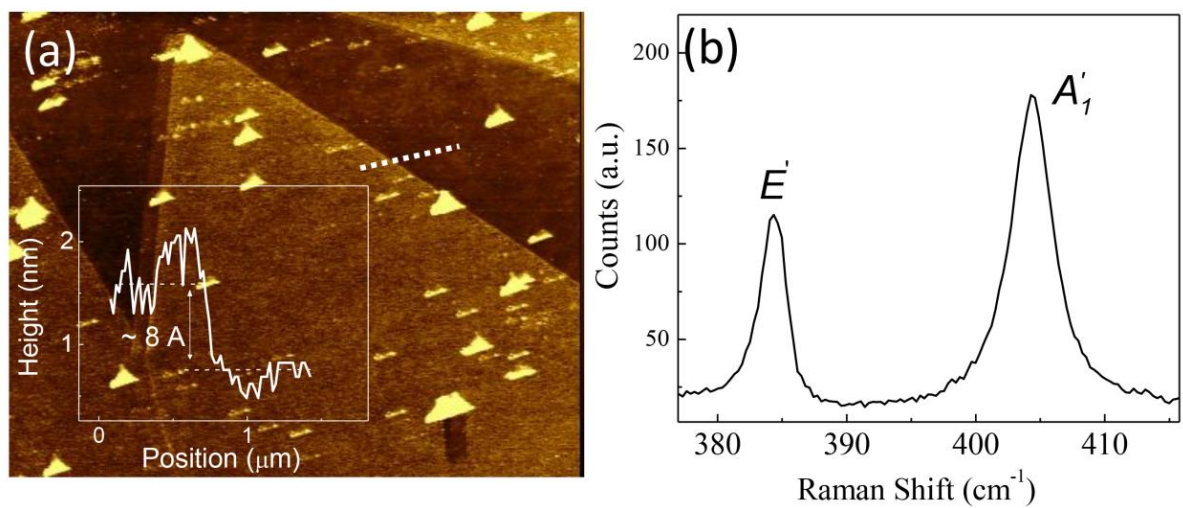

**Figure S3:** (a) AFM image of a CVD grown sample. The height profile along the black dash is shown in the inset. (b) Raman spectroscopy of a CVD sample.

## S4: Optical image of CVD grown 1L-MoS$_2$

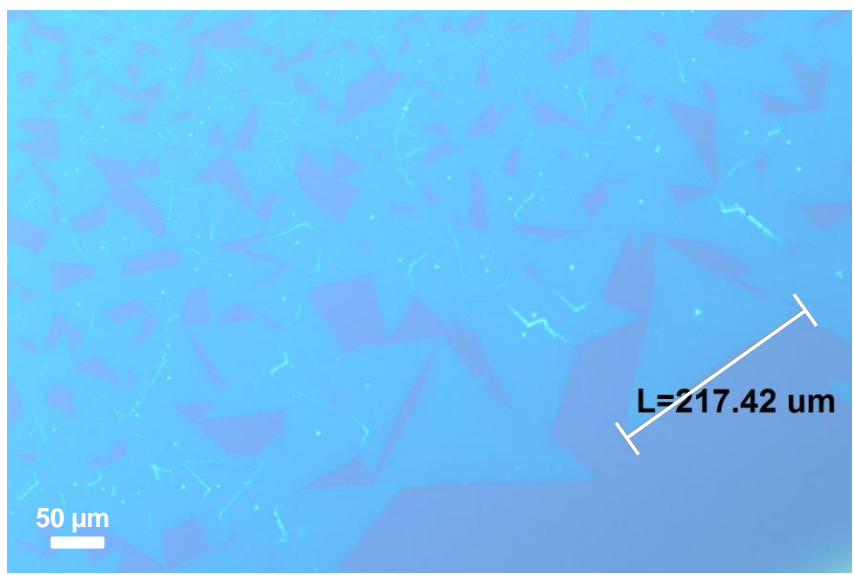

**Figure S4:** The optical image of CVD grown 1L-MoS2.

## S5: Calculation of Local electrostatic potential and charge distribution

Fig. S5 shows the calculated electrostatic potential along the ribbon. The asymmetric potential, which is lower at the Mo-edge, where the oxygen dimer absorbs, and lifted at the S-edge, as compared to the symmetric curve of the pristine MoS$_2$ ribbon, is caused by the oxygen atoms. The downshifted local electrostatic potential agrees with the lower valence band position close to the Mo-edge (Figure 4 in the main text). We believe that the excess electrons at the Mo-edge and the oxygen atoms cause a surface dipole pointing towards the nanoribbon leading to the lower local electrostatic potential, or correspondingly a higher local work function.

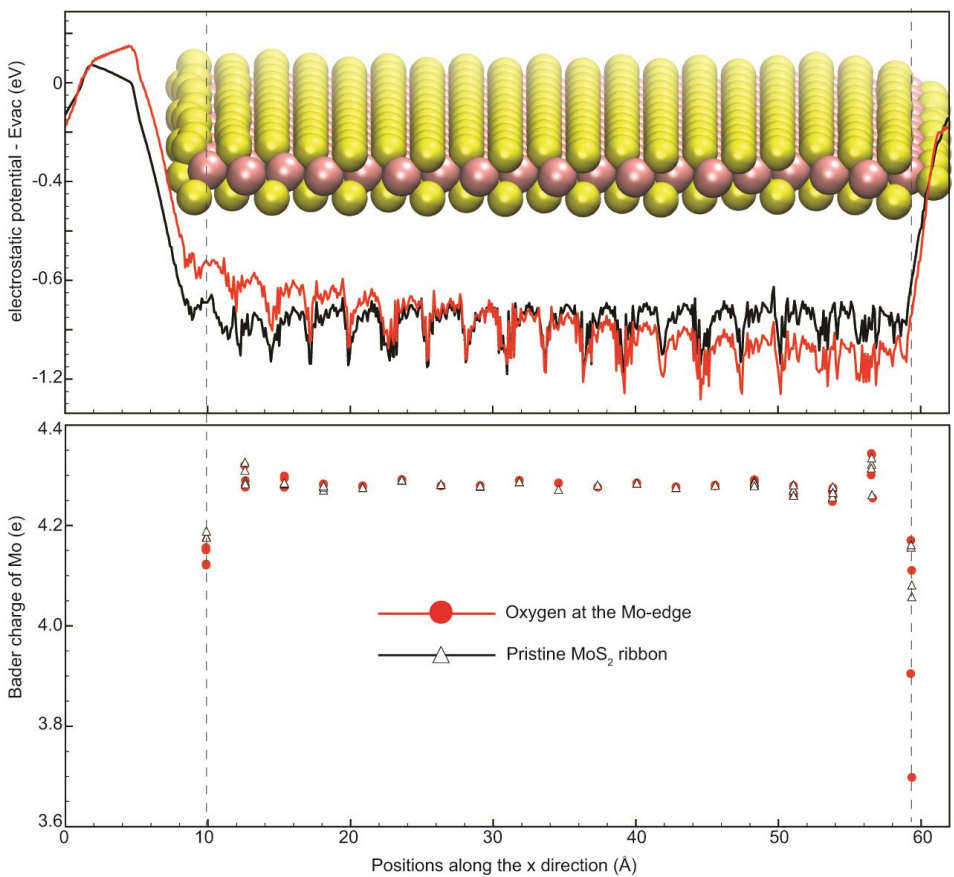

**Figure S5:** Calculated local electrostatic potential and Bader charge distribution with (red) and without (black) oxygen at the Mo-edge of the $MoS_2$ nanoribbon. (top) The averaged electrostatic potential of the *y*-plane, (perpendicular to the paper) and along the infinite ribbon direction, plotted as a function of the position along the *x*-direction. Vacuum level is set to zero. (bottom) the Bader charge plotted as a function of the *x* coordinate of each Mo atom. The two dashed lines indicate the edge positions.

Moreover, Bader charge[1] in Fig. S5 (bottom panel) shows that the Mo atoms at the Mo-edge have less electrons when an oxygen dimer replaces an S atom. It should be noted that Mo has 6 valence electrons. The Bader charge suggests that most Mo atoms are $Mo^{2+}$, and oxygen further withdraws electrons due to its higher electronegativity leading to a higher oxidation state of the Mo atom at the Mo-edge. The metallic nature of the Mo-edge and the excess electrons at the oxygen helps to screen the coulombic interaction in the exciton.

References:

(1)  Henkelman, G.; Arnaldsson, A.; Jónsson, H.: A fast and robust algorithm for Bader decomposition of charge density. *Computational Materials Science* 2006, *36*, 354-360.